\documentstyle{article}

\def\lsim{\mathrel{\vcenter{\hbox{$<$}\nointerlineskip\hbox{$\sim$}}}}
\newcommand{\be}{\begin{equation}}
\newcommand{\ee}{\end{equation}}
\newcommand{\ba}{\begin{eqnarray}}
\newcommand{\ea}{\end{eqnarray}}
\newcommand{\nn}{\nonumber}

\begin{document}
\vspace*{-1in}
\renewcommand{\thefootnote}{\fnsymbol{footnote}}
\begin{flushright}
\texttt{hep-ph/???} 
\end{flushright}
\vskip 5pt
\begin{center}
{\Large{\bf Effects of quantum space time foam in the neutrino sector}}
\vskip 25pt
{\sf H.V. Klapdor-Kleingrothaus $^{1}$},
{\sf H.  P\"{a}s $^{1}$},
{\sf U. Sarkar $^{2}$}
\vskip 10pt
$^1$ {\small Max-Planck-Institut f\"{u}r Kernphysik, P.O. Box 103980,
D-69029 Heidelberg, Germany} \\
$^2$ {\small Physical Research Laboratory, Ahmedabad, 380 009, India}

\vskip 20pt

{\bf Abstract}
\end{center}

\begin{quotation}
{\small 
We discuss violations of CPT and quantum mechanics due to interactions 
of neutrinos with space-time quantum foam. Neutrinoless double 
beta decay and oscillations of neutrinos from astrophysical sources 
(supernovae, active galactic nuclei) 
are analysed. It is found that the propagation distance 
is the crucial quantity entering any bounds on EHNS parameters. Thus, while
the bounds from neutrinoless double beta decay are not significant,
the data of the
supernova 1987a imply a bound being several orders of magnitude more 
stringent than the ones known from the literature.
Even more stringent limits may be obtained from the investigation of  
neutrino oscillations from active galactic nuclei sources, which have an 
impressive potential for the search of quantum foam interactions in the 
neutrino sector. 
}
\end{quotation}

\vskip 20pt  

\setcounter{footnote}{0}
\renewcommand{\thefootnote}{\arabic{footnote}}

\section{Introduction}
While in the context of local quantum field theories CPT has to be conserved, 
CPT violating effects may show up in the framework of quantum gravity.
As an example, Hawking radiation of black holes can be understood as a pair
creation process near the event horizon, with one particle falling into the 
black hole and the other one escaping. Since with the particle falling into 
the black hole some phase information of the quantum state is lost, the
thermic final state is a mixed state rather than a pure one. As Hawking has
pointed out \cite{hawk}, such an evolution of a  pure state into a mixed 
state violates
the laws of conventional quantum mechanics (QMV). 
If the space time possesses 
a foamy structure at the Planck scale, 
including the creation and annihilation 
of black holes with Planck radius and Planck lifetime, such effects also may
influence microscopical processes in the vacuum \cite{hawk2}.
In the following Page \cite{page} showed that such processes violate also CPT
and the 
possibility of 
experimental tests in the $K_0-\bar{K_0}$ sector was discussed by Eberhard 
\cite{eber}. Ellis, Hagellin, Nanopoulos and Srednicki independently 
developed an evolution equation formalism in the space of density matrices
\cite{EHNS} containing three CPT violating (EHNS)
parameters $\alpha,\beta,\gamma$
which have a dimension of mass and which might be expected to be of order 
$m_K^2/M_{Pl} \sim 10^{-20}$ GeV in the Kaon sector. 
Recently the topic has been reconsidered by Ellis, Mavromatos and Nanopoulos
\cite{EMN}
and Huet and Peskin \cite{huet}. 
CPT violating processes in the neutrino sector have been discussed for the 
first time by Liu et al. \cite{liu} and in the following in \cite{liu2}, where
neutrino oscillations due to CPT violation has been discussed as a solution to 
the solar neutrino problem. Recently another paper \cite{lisi}
explored the possibility of explaining the atmospheric neutrino
anomaly with quantum foam effects and came to a negative conclusion. 
In this note we extend the discussion of quantum 
foam effects in the neutrino sector to the cases of neutrinoless double beta 
decay and oscillations of neutrinos from 
astrophysical sources, supernovae as well as active galactic nuclei.
New, extremely stringent bounds are found improving constraints found in the 
literature by several orders of magnitude.

\section{Density matrix formalism}

For mixed states it is useful to work in the framework of the density
matrix formalism, following the methodology as presented in ref.
\cite{liu}. We start with the Schr\"odinger equation for the density 
matrix, 
\be
i \frac{d}{dt} \rho = [H,\rho].
\ee
Here $\rho$ is the density matrix of the system, which can be 
expanded in the Pauli matrix basis,
\be
\rho=\rho^0 I + \rho^i \sigma^i,
\ee
where $I$ is the unity matrix and $\sigma^i$ are the Pauli matrices.
In \cite{liu2} a lepton number violating parametrization 
for the evolution equation of the components of the density matrix
has been assumed:

\be
\frac{d}{dt}\left(
\begin{array}{c} \rho^0\\ \rho^1\\ \rho^2\\  \rho^3
\end{array}
\right)=2 \left(
\begin{array}{cccc}
0 &0 & 0 & 0\\
0 & 0 & \Delta m^2/(4E) & 0\\
0 & -\Delta m^2/(4E) & -\alpha & -\beta\\
0 &0 &-\beta & -\gamma
\end{array}
\right)
\left(
\begin{array}{c} \rho^0\\ \rho^1\\ \rho^2\\  \rho^3
\end{array}
\right).
\ee
Here $\beta \ll \alpha, \gamma$ \cite{EHNS,huet,liu}. 
In \cite{liu,liu2} also an alternative, lepton-number conserving 
parametrization has been discussed. However this parametrization will not 
influence neither the double beta decay observable nor the oscillation 
probability in the asymptotics of large propagation distances compared to the
standard case of neutrino masses \cite{liu2}. Thus we will concentrate on the
lepton-number violating case in the following. It should be mentioned however
that a full analysis of the generalized dynamics requires 
six parameters \cite{BF}.

Moreover, 
it should be stressed that this non-relativistic ansatz may not be suitable 
to describe ultrarelativisic particles such as neutrinos. However, while the 
covariant treatment of open quantum systems is still an unsolved problem, 
the density matrix ansatz has been successfully used in previous works
to derive the ``standard'' mass mechanism neutrino oscillation probability
also, see \cite{lisi}. 
Thus, while future works should improve the present ansatz, 
this approach seems to be suitable to provide at least a possibility for the
comparison of the
sensitivity of different experiments and a rough estimation for the order 
of magnitude of the obtained bounds.

\section{Neutrinoless double beta decay}
Neutrinoless double beta decay is one of the most sensitive tools in neutrino 
physics. It corresponds to two single beta decays occuring simultaneous 
in one nucleus, with a virtual neutrino propagating between the vertices.
Important impact of this process has been derived on the reconstruction
of the neutrino mass spectrum, physics beyond the standard model as well as
more exotic phenomena such as violations of the equivalence principle or 
Lorentz invariance (for an overview see \cite{kp,kps}). 
In the following we will study the potential of 
neutrinoless double beta decay for searches for CPT violations due to quantum 
foam interactions in the neutrino sector.
The observable measured in neutrinoless double beta decay is the $ee$ entry
of the neutrino mass matrix in the flavor space,
\be
m_{ee}= \bar{m}-\frac{\delta m}{2} cos (2 \theta)
\ee
in a two neutrino scenario with $\bar{m}=(m_1+m_2)/2$ and 
$\delta m= (m_2 - m_1)$ and $m_{1,2}$ being the mass eigenstates.
This quantity will be modified in the presence of QMV. 
The recent experimental constraint is
$m_{ee}<0.3$ eV, obtained from the Heidelberg--Moscow experiment searching for 
double beta decays of $^{76}$Ge \cite{baudis}. The GENIUS project will
be sensitive to $m_{ee}=10^{-2}-10^{-3}$ eV \cite{genius}.
In the density matrix formalism the double beta decay observable can be 
expressed as follows:
\ba
Tr ( \rho_{\nu_e} \cal{O})&=& Tr \left(
\begin{array}{cc}
\rho_0 + \rho_3 & \rho_1 - i \rho_2\\
\rho_1 + i \rho_2 & \rho_0 + \rho_3 \\
\end{array}
\right).
\left(
\begin{array}{cc}
m_1 & 0\\
0 & m_2 \\
\end{array}
\right)
\nn \\
&=& (m_1 + m_2) \rho_0 + (m_2 - m_1) \rho_3.
\ea

The propagation time of the neutrino 
\be
t=\frac{1}{4 \pi \Delta E} \simeq 6 \cdot 10^{-24} s
\ee
can be estimated by taking its energy to be of the size of the nuclear
Fermi momentum $p_F \simeq 100$ MeV for $^{76}$Ge.
Assuming $\beta \ll \alpha, \gamma$, eq. (5) yields
\ba
\frac{d}{dt}\rho_0&=&0 \\ 
\frac{d}{dt}\rho_3&=&-2 \gamma \rho_3 
\ea
and thus, using eq. (4)
\ba
\rho_0 &=& \frac{1}{2} \\
\rho_3 &=& e^{-2 \gamma t} \frac{cos(2 \theta)}{2}. 
\ea
This implies
\be
m_{ee}^{QMV}= \bar{m} + e^{-2 \gamma t} \frac{\Delta m}{2} \cos 2 \theta. 
\ee
Due to the tiny propagation time (6)   no significant variation
of the double beta decay observable is obtained. However, from this
analysis we realize that the distance plays a crucial role in
constraining the QMV parameters, so we shall consider the bounds on
the neutrino oscillation probability where neutrinos are propagating over  
large distances.

\section{Oscillations of neutrinos from astrophysical sources}

In the following we study the effect of quantum mechanics violation  
in neutrino
oscillations from astrophysical sources.  The most distant sources that have 
been discussed in the context of neutrino oscillations are supernovae (SN) and 
active galactic nuclei (AGN). While astrophysical sources have been discussed 
in the context of QMV effects on life time measurements \cite{astro,astro2}, 
they have not been considered for the case of QMV induced
neutrino oscillations so far.

For the neutrino oscillation case we get the survival and disappearance
oscillation probabilities \cite{liu,liu2}
\ba
P(\nu_x \rightarrow \nu_{x})&=&Tr[\rho_{\nu_x}(t)\rho_{\nu_x}] \\
P(\nu_x \rightarrow \nu_{x'})&=&Tr[\rho_{\nu_x}(t)\rho_{\nu_{x'}}],
\ea
respectively. Here the density matrices can be parametrized as
\ba
\rho_{\nu_x}&=&
\left(
\begin{array}{cc}
\cos^2 \theta & \cos \theta \sin \theta\\
\cos \theta \sin \theta & \sin^2 \theta \\
\end{array}
\right), \\
\rho_{\nu_{x'}}&=&
\left(
\begin{array}{cc}
\sin^2 \theta & -\cos \theta \sin \theta\\
-\cos \theta \sin \theta & \cos^2 \theta \\
\end{array}
\right). 
\ea
As initial condition we assume
\begin{equation}
\rho(t=0)=\rho(\nu_e)
\end{equation}
and thus \cite{liu,liu2}:
\ba
\rho_0&=&\frac{1}{2}\\
\rho_1&=&\frac{1}{2}\sin(2 \theta)\\
\rho_2&=&0\\
\rho_3&=&\frac{1}{2} \cos(2 \theta).
\ea
The interesting observable is the oscillation propability
\begin{equation}
P^{QMV}_{ \nu_{x} \to \nu_{x'} } 
= Tr[\rho(t) \rho_x]=\frac{1}{2}-\frac{1}{2} e^{-\gamma L}\cos^2 2 \theta
-\frac{1}{2} e^{-\alpha L} \sin^2 2 \theta \cos (\frac{\Delta m^2}{2 E_\nu}L),
\end{equation}
where $\beta \ll \alpha, \gamma$ has been assumed.
For the n-flavour case the oscillation probability 
for large propagation distances is given by \cite{liu2},
\begin{equation}
P^{QMV}_{ \nu_{x} \to \nu_{x'} } 
= {1 \over n} - {1 \over n} e^{- \gamma L},
\end{equation}
where L is the propagation distance of the neutrinos. 

This QMV oscillation probability can easily be distinguished from the
asymptotics of the ``standard'' mass induced oscillation probability:
\be
P^{mass}_{ \nu_{x} \to \nu_{x'} } 
= \frac{\sin^2 2 \theta}{2}.
\ee
The quantity $P^{mass}$ is fixed experimentally to 
$P^{mass}_{ \nu_{\mu} \to \nu_{\tau} } \simeq 0.5$
due to the maximal mixing in atmospheric neutrinos \cite{superk}
and $P^{mass}_{ \nu_e \to \nu_{\tau} } \lsim 0.05$ 
due to the CHOOZ bound \cite{chooz}.

{\it Supernovae 1987a}: 
In supernovae strong neutrino oscillations will significantly 
distort the $\nu_e$ spectra at the earth, since the $\nu_e$ will aquire
the spectra of the more energetic $\nu_{\mu}$ and $\nu_{\tau}$.
The distance is very large.
As a result, the condition that QMV should satisfy the bound on the 
oscillation probability gives a very strong bound. In the case
of supernova 1987a, 
$L \sim 50 Mpc
\sim 7 \cdot 10^{39} GeV$, so that the observed constraint on the oscillation 
probability \cite{smirnov}
$ P^{exp}_{ \nu_e \to \nu_{\mu,\tau}} < 0.2$ is satisfied for the three 
neutrino case when 
\begin{equation}
\gamma < {0.6 \over L} \sim 10^{-40} GeV .
\end{equation}
We assumed here that $P^{exp}$ is the accuracy with which the
deviations from the
asymptotics $1/n=1/3$ can be measured. Due to the unknown energy 
dependence of the EHNS parameters and the Lorentz non-invariant ansatz
it is difficult to compare these bounds
with the bound coming from K-physics. Following \cite{liu} we 
assume $\gamma$ to be of the order $E_{\nu}^2/M_{Pl}$
and
scale the obtained bound by the neutrino energy to the kaon mass squared, 
\be
\gamma_{\nu} \propto \frac{E_{\nu}^2}{m_K^2},
\ee
implying  $ \gamma_K < 10^{-37} GeV$, which is 
an improvement of
about 16 orders of magnitude. 
This disfavors strongly any solution of the solar or
atmospheric neutrino problem by lepton number violating
QMV effects. If one assumes that the 
same QMV parametrization is valid for the $K-$system, then any
observational possibility in the $K-$system will also be excluded by
the present constraints from the supernovae analysis.
A relativistic treatment of the problem can modify this bound to some extent,
but it is most unlikely that the modification is by several orders of 
magnitude.

{\it Active Galactic Nuclei (AGN)} :
AGN can be intense sources of high energy neutrinos ($E_{\nu} \sim {\cal O}$
(1 PeV)) \cite{pakv}.
According to representative models the flux of these neutrinos is
flavor dependent and the $\nu_{\tau}$ flux is reduced by at least two orders 
of magnitude compared to the $\nu_e$, $\nu_{\mu}$ fluxes. An unique appearance 
signal of high energy $\nu_{\tau}$ neutrinos can be a double bang signal of
the produced $\tau$ leptons: The first bang originates from the CC interaction
of the $\tau$ neutrino and the second one from the hadronic decay of the 
$\tau$ lepton. Deep underwater or ice neutrino detectors have been estimated
to be sensitive on neutrino oscillation probabilities of \cite{pakv,smivep}
\be
P^{exp}_{ \nu_{e,\mu} \to \nu_\tau } < 5 \times 10^{-3}.
\ee
Since the QMV effects become strong for large distances
and higher energies, 
it is likely that when we have data from the active galactic nuclei on
neutrino oscillations, these bounds will be modified by several orders 
of magnitude. Considering the distance to be $L \sim 100 Mpc$ and the
average energy of the neutrinos to be around 1 PeV, a bound on
the neutrino oscillation of $ P_{ \nu_e \to \nu_\mu } < 5 \times 10^{-3}$
will imply
a corresponding bound on the QMV parameter 
\be
\gamma_{\nu} < 10^{-42} GeV.
\ee
Translation to the kaon mass scale yields
\begin{equation}
\gamma_K < 10^{-55} GeV, 
\end{equation}
which would imply the by far strongest bound on QMV parameters.
This will provide a decisive test for any contribution of lepton number 
violating QMV effects
in the neutrino sector.

\section{Conclusions}

We studied the effects of violation of quantum mechanics 
due to quantum space time foam interactions in neutrino experiments. While
the non-observation of  neutrinoless double beta decay does not give
any significant constraint, the supernova 1987a implies a constraint being
16 orders of magnitude more stringent than the bounds known from the 
literature. This
disfavors strongly any possibility of observable effects of lepton number 
violating
QMV in any other
experiments. The non-observation of QMV induced neutrino 
oscillations from active galactic nuclei 
will be able to improve this bound by many orders 
of magnitude. 
While the chosen non-relativistic ansatz might not be totally suitable for
neutrinos, it should be at least useful to compare the sensitivity of 
different neutrino sources. Moreover the bounds obtained are that stringent,
that, even in view of this ambiguity, they should be considered as the most 
restrictive ones. 

\section*{Acknowledgements}
We thank J. Ellis, E. Lisi, S. Pakvasa, A.Y. Smirnov, the referee 
and especially N.E. Mavromatos
for comments and useful discussions.


\begin{thebibliography}{99}

\bibitem{hawk}
S.W. Hawking, Nature 248, (1974) 30

\bibitem{hawk2}
S.W. Hawking, Commun. Math. Phys. 43 (1975) 199

\bibitem{page}
D.N. Page, Gen. Relativ. Gravit. 14 (1982) 1; 
Phys. Rev. Lett. 44 (1980) 301 

\bibitem{eber}
P.H. Eberhard, CERN report 72-1 (1972) unpubl.; W.C. Carithers, J.H. 
Christenson, P.H. Eberhard, D.R. Nygren, T. Modis, T.P. Pun, E.L. Schwartz,
H. Sticker, Phys. Rev. D 14 (1976) 290

\bibitem{EHNS}
J. Ellis, J.S. Hagellin, D.V. Nanopoulos, M. Srednicki, Nucl. Phys. B 241 
(1984) 381

\bibitem{EMN}
J. Ellis, N.E. Mavromatos, D.V. Nanopoulos, Phys. Lett. B 293 (1992) 193

\bibitem{huet} 
P. Huet, M.E. Peskin,  Nucl.Phys. B434 (1995) 3-38

\bibitem{liu}
Y. Liu, L. Hu, M.-L. Ge, Phys. Rev. D 56 (1997) 6648

\bibitem{liu2}
Y. Liu, J.-L. Chen, M.-L. Ge, J. Phys. G 24 (1998) 2289;
F. Ma, H. Hu, hep-ph/9805391;
C.-H. Chang, W.-S. Dai, X.-Q. Li, Y. Liu, F.-C. Ma, Z.-J. Tao,
Phys. Rev. D 60 (1999) 033006

\bibitem{BF} 
F. Benatti, R. Floreanini, JHEP 02 (2000) 032

\bibitem{lisi}
E. Lisi, A. Marrone, D. Montanino, hep-ph/0002053

\bibitem{kp}
H.V. Klapdor--Kleingrothaus, H. P\"as, hep-ph/0002109, in Proc. 
{\it Cosmo '99, Trieste, Italy} 

\bibitem{kps}
H.V. Klapdor--Kleingrothaus, H. P\"as, A.Y. Smirnov,
hep-ph/0003219

\bibitem{baudis}
HEIDELBERG--MOSCOW collab., Phys. Rev. Lett. 83 (1999) 41; priv. comm.

\bibitem{genius}
H.V. Klapdor--Kleingrothaus, Proc. {\it Beyond the Desert '97};
J. Hellmig, H.V. Klapdor--Kleingrothaus, Z. Phys. A 359 (1997) 351;
H.V. Klapdor--Kleingrothaus, M. Hirsch, Z. Phys. A 359 (1997) 361;
H.V. Klapdor--Kleingrothaus, J. Hellmig, M. Hirsch, J. Phys. G 24 (1998)
483;
H.V. Klapdor--Kleingrothaus, L. Baudis, G. Heusser, B. Majorovits, 
H. P\"as, hep-ph/9910205

\bibitem{astro}
J. Ellis, N.E. Mavromatos, D.V. Nanopoulos, G. Volkov,
gr-qc/9911055

\bibitem{astro2}
O. Bertolami, C.S. Carvalho, gr-qc/9912117


\bibitem{smirnov}
A.Y. Smirnov, D.N. Spergel, J.N. Bahcall, Phys. Rev. D 49 (1994) 1389

\bibitem{pakv}
J.G. Learned, S. Pakvasa, Astropart. Phys. 3 (1995) 267-274;
S. Pakvasa, hep-ph/9503369;  
T.J. Weiler, W.A. Simmons, S. Pakvasa, J.G. Learned, hep-ph/9411432

\bibitem{smivep}
H. Minakata, A.Y. Smirnov, Phys.Rev. D54 (1996) 3698-3705

\bibitem{superk}
Y. Fukuda et al. (SuperKamiokande Collab.), Phys. Lett. B 433 (1998) 9

\bibitem{chooz}
M. Apollonio et al. (CHOOZ collab.), hep-ex/9907037, Phys. Lett. B 466 (1999)
415-430


\end{thebibliography}
\end{document}